\documentclass[aip,amsmath,amssymb,graphicx,reprint]{revtex4-1}

\usepackage{graphicx}
\usepackage{dcolumn}
\usepackage{bm}

\usepackage{amsmath,braket}
\usepackage{mathtools}
\usepackage{color}
\usepackage{xcolor}
\usepackage{setspace}

\DeclareUnicodeCharacter{2212}{-}

\draft 

\begin{document}

\title{Tuning the mode-splitting of a semiconductor microcavity with uniaxial stress}

\author{Natasha Tomm}
\email{natasha.tomm@unibas.ch}
\affiliation{Department of Physics, University of Basel, Klingelbergstrasse 82, CH-4056 Basel, Switzerland}
\author{Alexander R.\ Korsch}
\affiliation{Lehrstuhl f\"{u}r Angewandte Festk\"{o}rperphysik, Ruhr-Universit\"{a}t Bochum, D-44780 Bochum, Germany}
\author{Alisa Javadi}
\affiliation{Department of Physics, University of Basel, Klingelbergstrasse 82, CH-4056 Basel, Switzerland}
\author{Daniel Najer}
\affiliation{Department of Physics, University of Basel, Klingelbergstrasse 82, CH-4056 Basel, Switzerland}
\author{R\"udiger Schott}
\affiliation{Lehrstuhl f\"{u}r Angewandte Festk\"{o}rperphysik, Ruhr-Universit\"{a}t Bochum, D-44780 Bochum, Germany}
\author{Sascha R.\ Valentin}
\affiliation{Lehrstuhl f\"{u}r Angewandte Festk\"{o}rperphysik, Ruhr-Universit\"{a}t Bochum, D-44780 Bochum, Germany}
\author{Andreas D.\ Wieck}
\affiliation{Lehrstuhl f\"{u}r Angewandte Festk\"{o}rperphysik, Ruhr-Universit\"{a}t Bochum, D-44780 Bochum, Germany}
\author{Arne Ludwig}
\affiliation{Lehrstuhl f\"{u}r Angewandte Festk\"{o}rperphysik, Ruhr-Universit\"{a}t Bochum, D-44780 Bochum, Germany}
\author{Richard J.\ Warburton}
\affiliation{Department of Physics, University of Basel, Klingelbergstrasse 82, CH-4056 Basel, Switzerland}

\date{\today}

\begin{abstract}
A splitting of the fundamental optical modes in micro/nano-cavities comprising semiconductor heterostructures is commonly observed. Given that this splitting plays an important role for the light-matter interaction and hence quantum technology applications, a method for controlling the mode-splitting is important. In this work we use an open microcavity composed of a ``bottom" semiconductor distributed Bragg reflector (DBR) incorporating an n-i-p heterostructure, paired with a ``top" curved dielectric DBR. We measure the mode-splitting as a function of wavelength across the stopband. We demonstrate a reversible {\it in-situ} technique to tune the mode-splitting by applying uniaxial stress to the semiconductor DBR. The method exploits the photoelastic effect of the semiconductor materials. We achieve a maximum tuning of $\sim$11 GHz. The stress applied to the heterostructure is determined by observing the photoluminescence of quantum dots embedded in the sample, converting a spectral shift to a stress via deformation potentials. A thorough study of the mode-splitting and its tuning across the stop-band leads to a quantitative understanding of the mechanism behind the results.
\end{abstract}

\maketitle

Semiconductor quantum dots (QDs) coupled to optical microcavities represent an important platform to advance quantum information technologies. Semiconductor QD-cavity platforms, such as micropillars, photonic crystals and open microcavities, have been successfully employed to achieve highly efficient single-photon sources \cite{Wang2019,Tomm2021}, a coherent light-matter interaction \citep{Najer2019}, generation of entangled photons \cite{Dousse2010,Liu2019}, and photon-photon switches \cite{Fushman2008}. Despite the history of successful cavity quantum-electrodynamics demonstrations in these systems, there are still partly unresolved technical questions that affect their performance. One such property is the almost ubiquitous observation that the fundamental cavity mode splits into two separate modes with linear, orthogonal polarizations. This lifting of the polarization degeneracy is desired and exploited in some cases, notably in efficient single-photon sources in order to avoid a 50\% loss of signal in cross-polarized collection schemes \cite{Wang2019,Tomm2021}. In this scenario, a QD trion is excited via one cavity mode, and photons are preferentially emitted into the other cavity mode. In other cases however, polarization degenerate cavity modes are desirable. This is typically the case in experiments relying on circularly polarized excitation schemes \cite{Lodahl2017}, for instance a single spin in a perpendicular magnetic field. Here, the linearly polarized cavity modes result in a reduced coupling to the quantum emitter \cite{Sollner2015}. It is not simple to control the bare mode-splitting precisely -- it can depend on the local inbuilt strain in the material, and on fabrication imperfections. For all these reasons, a way of selectively tuning and controlling the mode-splitting is of great interest.

The polarization splitting of a semiconductor microcavity's fundamental mode is the result of birefringence in the semiconductor between two orthogonal crystalline axes (which are themselves orthogonal to the optical axis). In zinc-blende type crystals there is {\it a priori} no intrinsic birefringence. Birefringence can be created however, often unintentionally, via two mechanisms. First, in heterostructures incorporating a diode or Schottky structure, the in-built electric field along the $z$ direction (growth axis) breaks the inversion symmetry of the crystal and birefringence in the $x$-$y$ plane arises via the linear electro optic effect \cite{vanExter1997}. Secondly, a uniaxial stress in the $x$-$y$ plane, induced by microscopic imperfections in the heterostructure or post-growth processing, induces birefringence via the photoelastic effect \cite{Ziel1977,Raynolds1995}. Contrarily, a biaxial stress does not result in observable birefringence on account of the symmetry of the zinc-blende crystal.

\begin{figure}[b]
\includegraphics[width=\columnwidth]{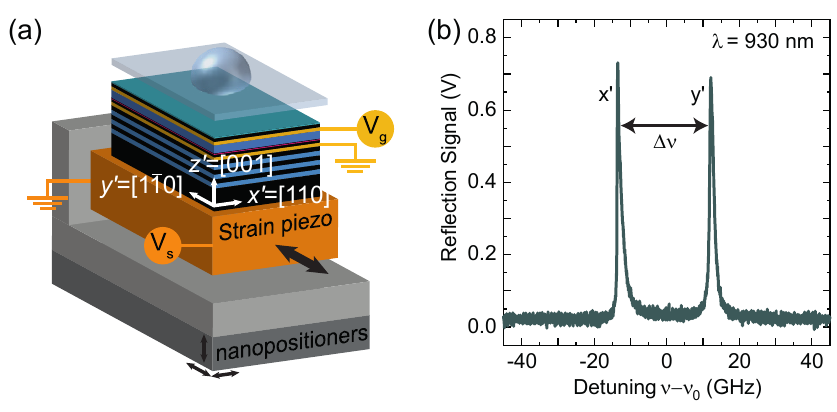}
\caption{\label{fig:setup}(a) Experimental setup depicting the microcavity composed of dielectric top mirror, and a semiconductor heterostructure, containing InAs QDs embedded in an n-i-p diode structure (applied voltage $V_{\rm g}$), and the bottom mirror. The sample is glued onto a piezostack (applied voltage $V_{\rm s}$), that stresses the sample along the $y'$ crystalline direction. The sample is positioned both laterally and vertically relative to the top mirror via nanopositioners. (b) Dark-field reflectivity scan across a cavity resonance: the fundamental mode is split into two linearly and orthogonally polarized modes. The microcavity axes are aligned with respect to the sample's crystalline axes $x'$ and $y'$. The $x'$-polarized ($y'$-polarized) mode is red (blue) detuned from the expected resonance $\nu_0$. The orientation of the cavity modes is experimentally determined by aligning the polarization of the probing laser light to each cavity mode in turn, and by observing the alignment to the sample.}%
\end{figure}

One can use the electrooptic and photoelastic effects to reverse or enhance the birefringence in semiconductor cavities, as previously demonstrated in monolithic structures \cite{Bonato2009,Frey2018,Gerhardt2020}. Here, we present a way of tuning the mode-splitting of an open microcavity by making use of the photoelastic effect, i.e.\ the control of the birefringence upon application of uniaxial stress. A change in mode-splitting of $\sim 11$\,GHz is achieved. Moreover, application of uniaxial stress to an open microcavity results in control not only of the mode-splitting in the microcavity but also the absolute emission frequency of an embedded QD \cite{Seidl2006,Zhai2020}. In this microcavity embodiment, the full stress is experienced by the entire heterostructure. This is not necessarily the case for monolithic systems. 

We employ an miniturized Fabry-Pérot cavity\cite{Barbour2011,Greuter2014,Najer2019,Tomm2021}. The bottom mirror is a 46-pair AlAs($\lambda/4$)/GaAs($\lambda/4$) semiconductor distributed Bragg reflector (DBR) grown on a $[001]$ GaAs substrate, where $\lambda$ refers to the wavelength of light in the material. The surface of the semiconductor heterostructure is passivated via an Al$_2$O$_3$ layer~\cite{Najer2021}. The top mirror is a 15-pair SiO$_2$($\lambda/4$)/Ta$_2$O$_5$($\lambda/4$), Ta$_2$O$_5$-terminated, dielectric DBR where the layers are deposited onto a microcrater in a silica substrate. The semiconductor heterostructure contains a layer of InAs QDs; the QDs themselves are embedded within an n-i-p heterostructure, allowing the QD charge to be controlled via a voltage ($V_g$) applied to the diode \cite{Najer2019,Tomm2021}. The sample  is tightly glued onto a piezostack (PSt 150/7x7/7 cryo, Piezomechanik GmbH, Munich), as depicted in Fig.\,\ref{fig:setup}(a). The $[1\overline{1}0]$ direction of the crystal aligns with the polarization axis of the pieozstack such that application of a voltage $V_{\rm s}$ to the piezostack induces a $[1\overline{1}0]$-stress in the semiconductor. The spring constant of the sample is small compared to that of the piezostack, $k_{\rm sample} \ll k_{\rm PZT}$, such that the extension of the piezo should be unaffected by the attached semiconductor. The piezo-sample assembly is free to move relative to the top mirror laterally, allowing different positions in the sample to be probed, and vertically, allowing a reflection spectrum of the microcavity to be recorded at fixed laser wavelength. We employ a cross-polarization confocal microscope \cite{KuhlmannRSI2013}, where an added half-wave plate (HWP) allows the probe laser's polarization to be aligned with one or the other polarized cavity mode. The sample's orientation relative to the microscope axes is known; the cleaved edges of the semiconductor sample along the $x'=[110]$ and $y'=[1\overline{1}0]$ crystalline axes coincide with the microscope orientation to within few degrees. All experiments were carried out at a temperature T = 4\,K.

The fundamental cavity mode is probed by measuring the reflectivity of a narrowband laser in a polarization dark-field modus. Two closely spaced modes are observed as shown in Fig.\,\ref{fig:setup}(b). When the HWP is set such that the probe laser's polarization is aligned to $x'$ ($y'$), only the red (blue) detuned resonance is probed. When the HWP is set such that the probe laser is aligned at $45^{\circ}$ to the $x'$ and $y'$ directions, both cavity modes can be seen (Fig.\,\ref{fig:setup}(b)). These are the characteristic features of a birefringence-induced mode-splitting. The fact that the axes of the cavity modes are consistently aligned with the cleaved edges of the sample implies that birefringence arises in the semiconductor heterostructure, and not in the top mirror. Should the origin of the birefringence lie in the top mirror, no link to the crystal axes of the semiconductor would be expected.

\begin{figure}[t!]
\includegraphics[width=\columnwidth]{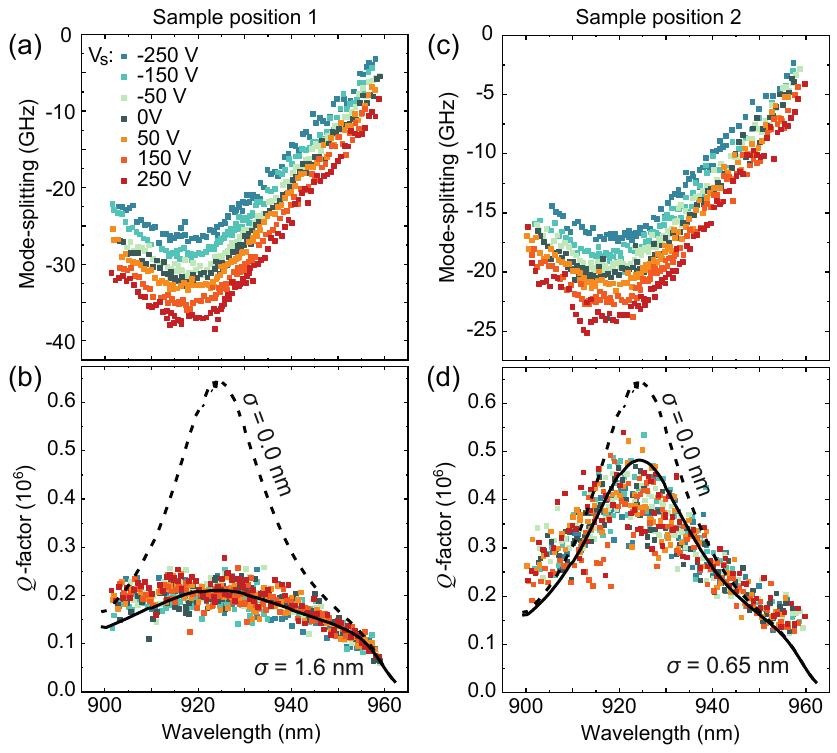}
\caption{\label{fig:maindata} Cavity mode-splitting $\Delta\nu=\nu_{x'}-\nu_{y'}$ as a function of probe wavelength and voltage $V_{\rm s}$ applied to the piezo at (a) position 1, and (c) position 2. The respective $\mathcal{Q}$-factors measured at these positions are shown in (b) and (d). Plots (b) and (d) show the modelled $\mathcal{Q}$-factor dispersion for this microcavity in the case without surface scattering $\sigma=0.0$\,nm at the semiconductor-vacuum interface (dashed line), and with surface scattering. The root-mean-square surface roughness is $\sigma=1.60$\,nm (b) and $\sigma=0.65$\,nm (d) for the two positions evaluated (solid lines).}
\end{figure}

The mode-splitting is defined by $\Delta \nu=\nu_{x'}-\nu_{y'}$ , where $\nu$ is the resonance frequency. It's important to note that the mode-splitting has a sign, negative in our case, meaning that the changes in refractive index along the $x'$ and $y'$ directions induce a red- and blue-shift, respectively, relative to the original resonance. The dynamic nature of the microcavity allows us to examine simultaneously the mode-splitting (Fig.\,\ref{fig:maindata}(a),(c)) and the $\mathcal{Q}$-factor across the microcavity's stop-band (Fig.\,\ref{fig:maindata}(b),(d)). Both the bare mode-splitting and the $\mathcal{Q}$ factor have a dependence on wavelength with maximum amplitude centered around $\lambda=918.7$\,nm, at the stop-band center.

We focus initially on the $\mathcal{Q}$-factors to demonstrate that we have a quantitative understanding of both the field confinement in the microcavity and the losses. We model the microcavity's stop-band and $\mathcal{Q}$-factor dependence on wavelength (Fig.\,\ref{fig:maindata}(b),(d) dashed and solid lines) using  a one-dimensional transfer-matrix simulation (Essential Macleod, Thin Film Center Inc.). In Fig.\,\ref{fig:maindata}(b),(d) the dashed lines depict the expected $\mathcal{Q}$-factor without any losses at the sample's surface. In practice, the measured $\mathcal{Q}$-factors are lower and this can be described very convincingly simply by including the effects of scattering at the Al$_2$O$_3$-vacuum interface~\cite{Najer2021}. The surface roughness was determined by comparing the experimental results and the theoretical model. We find that the maximum $\mathcal{Q}$-factor in this experiment depends on the exact lateral position, suggesting that the surface roughness changes across the sample~\cite{Najer2021}. A full wavelength dependence was acquired at two positions on the sample. A root-mean-square (rms) surface roughness of $\sigma = 1.60$\,nm ($\sigma = 0.65$\,nm) at position 1 (position 2) provide a very good description of the wavelength dependence of the $\mathcal{Q}$-factor. These surface roughnesses are consistent with characterization of the surface at room temperature with atomic force microscopy~\cite{Najer2021}. The residual small discrepancy between experimental and modelled curves probably arises from an imperfect knowledge of the exact layer thicknesses in the DBRs. 

\begin{figure}[t!]
\includegraphics[width=\columnwidth]{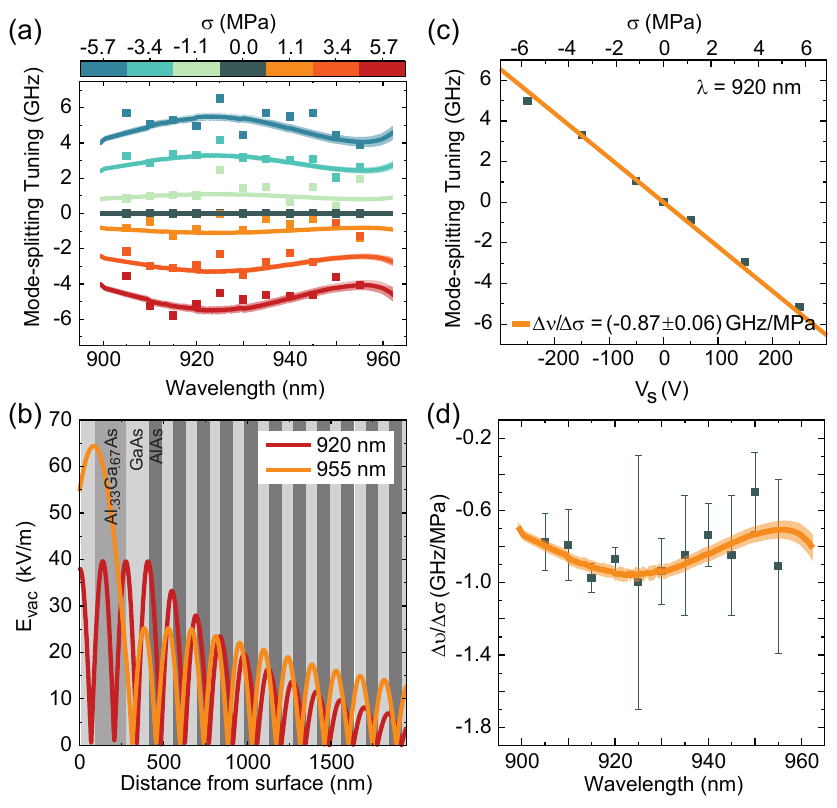}
\caption{\label{fig:Tuning_Vs} (a) Change in mode-splitting as a function of wavelength and applied uniaxial stress $\sigma$ from experimental data (dots) and theoretical model (solid lines). The model involves adjusting the refractive index of each layer in the heterostructure using Eqs.\,\ref{eq:DeltaSigmaDeltaVs} and \ref{eq:deltaN} and then calculating the exact resonance frequency in a one-dimensional transfer matrix simulation. The error bars in the model arise from uncertainties in the coefficients $q_{44}$ and in the calculation of $\delta E_{\rm gap}/\delta V_{\rm s}$. (b) Vacuum electric field distribution along the first few hundred nanometers below the sample's surface at wavelengths of 920\,nm and 955\,nm, indicating the dispersive influence of each layer's birefringence on the mode-splitting. (c) The mode-splitting tuning at $\lambda = 920$\,nm as a function of applied stress voltage $V_{\rm s}$ (and respective stress $\sigma$). A linear fit determines the tuning rate $\Delta\nu/\Delta\sigma = (-0.87\pm0.06)$\,GHz/MPa at this wavelength. (d) The tuning rate $\Delta\nu/\Delta\sigma$ as a function of wavelength across the the entire stop-band: experiment (black symbols); model (orange line).}
\end{figure}

We turn now to the behaviour on applying a uniaxial stress. We focus on position 1. Upon application of a voltage up to $V_{\rm s} = \pm 250$ V, the piezostack expands and contracts, thereby stressing the sample uniaxially along the $y'$ direction. The mode-splitting responds to the applied stress. A maximum mode-splitting tuning of approximately 11\,GHz (45\,$\mu$eV) is achieved at the exact wavelength where $\left|\nu_{x'}-\nu_{y'}\right|$ is the largest, as can be observed in Fig.\,\ref{fig:maindata}(a). The tuning leaves the $\mathcal{Q}$-factor unaltered (Fig.\,\ref{fig:maindata}(b)) indicating that the applied stress has no effect on the loss mechanisms in these high-$\mathcal{Q}$-factor cavities. The mode-splitting $\Delta \nu$ is a linear function of $V_{\rm s}$ (Fig.\,\ref{fig:Tuning_Vs}(b)); the response $\Delta \nu/\Delta V_{\rm s}$ is slightly smaller in magnitude at the edges of the stopband with respect to the stopband center (Fig.\,\ref{fig:Tuning_Vs}(d)).

\begin{figure}[b!]
\includegraphics[width=\columnwidth]{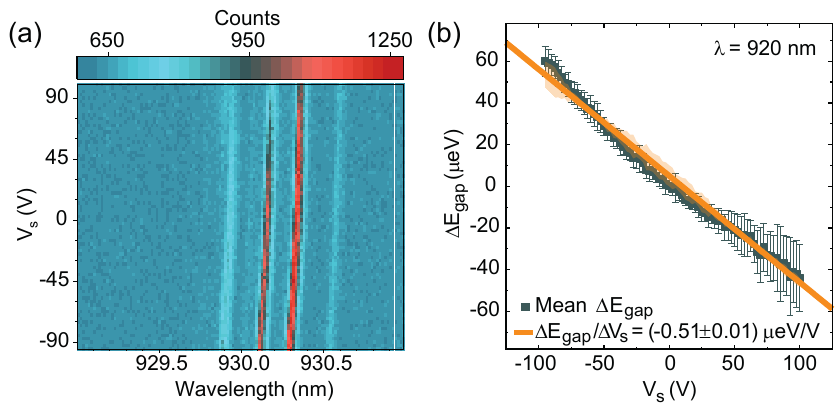}
\caption{\label{fig:deltaE_deltaVs} (a) Shift of the photoluminescence signal of a QD in the sample as a function of applied uniaxial stress. (b) Mean shift in bandgap energy as a function of voltage applied to the piezostack (V$_s$), as measured on 20 different excitonic lines in 10 QDs. A linear fit gives $\delta E_{\rm gap}/\delta V_{\rm s}= (-0.51 \pm 0.01)\mu$eV/V.}
\end{figure}

We now attempt to understand quantitatively the stress-induced change in mode-splitting. A crucial step is to determine the exact uniaxial stress applied. The extension per Volt of the piezostack depends strongly on temperature and unfortunately we do not know its exact value at T=4\,K. We do not have an external stress gauge in the experiment. Instead, we determine the applied stress by measuring the frequency-shift of the photoluminescence from the QDs embedded in the sample \cite{Seidl2006}. This has the advantage of determining the stress experienced by the heterostructure itself, exactly the stress which induces the birefringence. We determine the mean bandgap shift as a function of applied voltage $V_{\rm s}$ by observing the photoluminescence signal of 20 different excitonic lines in 10 QDs in the sample, as depicted in Fig.\,\ref{fig:deltaE_deltaVs}(a), and find $\delta E_{\rm gap}/\delta V_{\rm s}= (-0.51\pm0.01)\, \mu$eV/V equivalently $(-0.123\pm0.002)$\,GHz/V (Fig\,\ref{fig:deltaE_deltaVs}(b)), a value comparable to a previously achieved\cite{Seidl2006} tuning of $-0.82$\,$\mu$eV/V. The dominant effect of uniaxial stress on the emission frequency of the QDs is to induce a change in the bandgap of the host semiconductor GaAs \cite{Bhargava1967,Pollak1968,Higginbotham1969}, described by $\delta E_{\rm gap}/\delta \sigma$. The influence of uniaxial stress on the bandgap can be derived from the material's deformation potentials to be $\delta E_{\rm gap}/\delta \sigma = -22.2$ $\mu$eV/MPa, under the assumption that the valence state is pure heavy-hole. A detailed calculation is presented in the Appendix. Finally, from
\begin{equation}
\label{eq:DeltaSigmaDeltaVs}
\Delta \sigma/\Delta V_{\rm s} = \frac{\delta E_{\rm gap}/\delta V_{\rm s}}{\delta E_{\rm gap}/\delta \sigma}
\end{equation}
we infer $\frac{\Delta \sigma}{\Delta V_{\rm s}} = (22.97 \pm 0.45)$\,kPa/V, from which we are able to deduce the amount of stress applied to the sample $\sigma=\frac{\Delta \sigma}{\Delta V_{\rm s}} \, V_{\rm s}$.

The next step is to calculate the birefringence in each layer in the heterostructure. Stress-induced transformations to the dielectric function of a crystal are quantified by the so-called piezobirefringent tensor \cite{Nye1957,Higginbotham1969,Levine1992,Raynolds1995} $q_{ijkl}$. Due to the symmetry of zinc-blende crystals \cite{Nye1957,Raynolds1995}, and our system of coordinates $x'=[110]$, $y'=[1\overline{1}0]$, $z'=z=[001]$, the induced birefringence $\Delta n / n_0=\left(n_{x'}-n_{y'}\right)/n_0$ on stressing a semiconductor along $x'$ by an amount $\sigma$ is given by
\begin{equation}
\label{eq:deltaN}
    \frac{\Delta n}{n_0} = -\frac{n^{2}_{0}}{2} \cdot q_{44} \cdot \sigma,
\end{equation}
\noindent where $n_0$ is the bare refractive index of the particular material, and $q_{44}$ is a material parameter, $q_{44}=p_{44} \cdot S_{44}$, where $p_{44}$ is an element of the photoelastic tensor and $S_{44}$ an element of the compliance tensor. See Appendix for complete derivation.

Given that the sample is composed of layers of three different semiconductor materials (GaAs, Al$_{.33}$Ga$_{.67}$As and AlAs), the influence of uniaxial stress in each layer must be considered. The coefficients $q_{44}$ for GaAs, Al$_{.33}$Ga$_{.67}$As and AlAs at low temperature T = 4\,K are estimated (see Appendix for details) from literature room-temperature values \cite{Adachi1985} and found to be $q_{44} = \left(-7.4\pm1.2\right)\cdot 10^{-13}$\,m$^2$/N, $q_{44} = \left(-7.9\pm0.3\right)\cdot 10^{-13}$\,m$^2$/N and $q_{44} = \left(-1.64\pm0.02\right)\cdot 10^{-13}$\,m$^2$/N respectively.

Finally, we determine the mode-splitting by calculating the exact mode frequency for each polarization separately, including the subtle changes to the refractive indexes in the one-dimensional transfer-matrix simulation. Specifically, we use Eq.\,\ref{eq:deltaN} to calculate the induced birefringence $\Delta n$ in each layer of the heterostructure upon application of uniaxial stress $\sigma$, which is itself calculated with Eq.\,\ref{eq:DeltaSigmaDeltaVs}. For $V_{\rm s}=250$\,V, $\sigma = 5.74$\,MPa, the induced relative birefringence $\Delta n/n_0$ is as small as 26\,ppm in GaAs (25\,ppm in Al$_{.33}$Ga$_{.67}$As, 4\,ppm in AlAs). The stress-tuning of the mode-splitting is shown in Fig.\,\ref{fig:Tuning_Vs}(a) (solid lines) for each applied stress voltage $V_{\rm s}$ as a function of wavelength (spanning the stop-band). The results can be directly compared to the experimental results (symbols). Evidently from Fig.\,\ref{fig:Tuning_Vs}(a), the amount of tuning itself presents a dispersion, i.e. it depends on wavelength. The calculation captures this detail precisely and explains it: subtle shifts in the standing wave in the microcavity change the net birefringence as each layer of the heterostructure does not contribute equally. Figure\,\ref{fig:Tuning_Vs}(b) illustrates this point by showing the vacuum electric field as a function of distance from the sample's surface at a wavelength close to the stop-band center, at 920\,nm, and at a wavelength far away, at 955\,nm. As a consequence, the mode-splitting tunes linearly with stress, as depicted in Fig.\,\ref{fig:Tuning_Vs}(c) for $\lambda = 920$\,nm, but with different slopes $\Delta\nu/\Delta\sigma$ across the stop-band (Fig.\,\ref{fig:Tuning_Vs}(d)). Across the entire spectral range examined here, experimental data (points) and model (solid lines) present excellent agreement. 

Our method proves to be an effective way of controlling the intrinsic polarization splitting of an open semiconductor microcavity by up to $\pm 5.5$ GHz. The mode-splitting can be tuned across the entire stop-band in a predictable, reversible manner. The present microcavity has a rather large intrinsic mode-splitting. Nevertheless, the tuning capability allows us to achieve near-degeneracy of the cavity modes at the high-wavelength end of the stop-band. For a microcavity with a lower intrinsic mode-splitting, it should be possible to eliminate the mode-splitting. Of relevance here is the fact that the intrinsic mode-splitting and the applied stress are aligned along the same axes. The applied stress induces a small birefringence, on the order of a few ppm, and does not influence the microcavity's $\mathcal{Q}$ factor. The slight emission shift of the QDs embedded in the heterostructure can be compensated for in the present setup simply by exploiting the spectral tunability of the microcavity: a resonance with the cavity mode is easily maintained.

Naturally, it is desirable to achieve higher degrees of mode-splitting tunability, either to attain perfect degeneracy or to separate fully the two modes. An optimized architecture of the sample holder could increase the tuning rate~\cite{Zhai2020} by a factor of 20. The incorporation of a back-gate would allow us to apply an electric field across the bottom mirror, thereby making use of the electrooptic effect \cite{Frey2018}. Employing these two methods simultaneously would grant an even higher degree of control of the birefringence. A quantitative understanding of the origin of the intrinsic mode-splitting remains to be attained. However, we note that the mode-splitting dispersion curve can be used as a diagnostic tool, to indicate in which layers of the heterostructure birefringence is strong. 

\section*{Appendix}

\subsection{Photoelastic effect: the effect of uniaxial stress}
The refractive index $n$ of a crystal can be described by the indicatrix \cite{Nye1957}, an ellipsoid in which the principal axes represent the components of the dielectric tensor,
\begin{equation}
\label{eq:Indicatrix}
\overline{\overline{B_{ij}}}=\varepsilon_0 \frac{\partial E_i}{\partial D_j} = \left(\frac{1}{n^2}\right),
\end{equation}
where $\varepsilon_0$ is the vacuum's electric permittivity, $E_i$ is the electric field component along direction $i$ and $D_j$ is the electric displacement field along $j$. An applied stress deforms the indicatrix components $\Delta \overline{\overline{B_{ij}}}$ via the operation
\begin{equation}
\label{eq:IndicatrixPhotoelastic}
\Delta \overline{\overline{B_{ij}}}= \overline{\overline{q_{ijkl}}} \cdot \overline{\overline{\sigma_{kl}}}.
\end{equation}
Here, $\overline{\overline{q_{ijkl}}}$ ($i,j,k,l = 1,2,3$) is the fourth-rank piezo-birefringent tensor; the stress $\overline{\overline{\sigma_{kl}}}$ ($k,l = 1,2,3$) is a second-rank tensor. From Eq.\,\ref{eq:Indicatrix} and Eq.\,\ref{eq:IndicatrixPhotoelastic}, it follows that the change in refractive index $\Delta \overline{\overline{n_{ij}}} = \overline{\overline{n_{ij}}} - n_0$ (where $n_0$ is the bare refractive index of the isotropic material) reads
\begin{equation}
\label{eq:DeltaN}
\Delta \overline{\overline{n_{ij}}} = -\frac{\Delta \overline{\overline{B_{ij}}}}{2} \cdot n^3_0.
\end{equation}

For zinc-blende type (cubic) crystals, symmetry simplifies the photoelastic tensor such that only three independent coefficients remain \cite{Raynolds1995}, namely $q_{1111}$, $q_{1122}$ and $q_{2323}$. A compressed notation can be adopted: $11\rightarrow1$, $22\rightarrow2$, $33\rightarrow3$, $23\rightarrow4$, $13\rightarrow5$, $12\rightarrow6$. In this way, the rank of the tensors is reduced and the expression in Eq.\,\ref{eq:IndicatrixPhotoelastic} becomes $\Delta \overline{B_{m}}= \overline{\overline{q_{mn}}} \cdot \overline{\sigma_{n}}$ ($m,n = 1,2,3,4,5,6$). In extended form,
\begin{equation}
\label{eq:IndicatrixPhotoelasticReduced}
\left[ {\begin{array}{c}
\Delta B_{1}\\
\Delta B_{2}\\
\Delta B_{3}\\
\Delta B_{4}\\
\Delta B_{5}\\
\Delta B_{6}\\
\end{array} }
\right]
=
\left[ {\begin{array}{cccccc}
q_{11} & q_{12} & q_{12} & 0 & 0 & 0\\
q_{12} & q_{11} & q_{12} & 0 & 0 & 0\\
q_{12} & q_{12} & q_{11} & 0 & 0 & 0\\
0 & 0 & 0 & q_{44} & 0 & 0\\
0 & 0 & 0 & 0 & q_{44} & 0\\
0 & 0 & 0 & 0 & 0 & q_{44}\\
\end{array} }
\right]
\left[ {\begin{array}{c}
\sigma_{1}\\
\sigma_{2}\\
\sigma_{3}\\
\sigma_{4}\\
\sigma_{5}\\
\sigma_{6}\\
\end{array} }
\right].
\end{equation}

We now apply these general results to our problem. In the experiment, both the stress and the birefringence are applied/probed in the ($x'$,$y'$,$z'$) system of coordinates. Therefore, a rotation in the frame of reference by $\pi/4$ around $z=z'$ is applied. We treat the canonical case of a stress applied along the $x'$ direction.

In the ($x'$,$y'$,$z'$) basis, the simplified stress tensor for a uniaxial stress along $x'$ is self-evidently $\overline{\sigma'}=\left[1\,0 \,0 \,0 \,0 \,0 \right]^\intercal$. We start by calculating $\overline{\sigma}$ in the usual basis ($x$,$y$,$z$) from $\overline{\sigma'}$. The general rotation matrix  for an arbitrary angle $\theta$ and with $\theta = \pi/4$ is
\begin{equation}
\label{eq:RotationMatrix}
\overline{\overline{R}} = 
\left[ {\begin{array}{rrr}
\cos\theta & \sin\theta &  0\\
-\sin\theta & \cos\theta & 0\\
0 & 0 & 1
\end{array} }
\right]
\stackrel{\theta=\pi/4}{=}
\left[ {\begin{array}{ccc}
\frac{1}{\sqrt{2}} & \frac{1}{\sqrt{2}} & 0\\
\frac{-1}{\sqrt{2}} & \frac{1}{\sqrt{2}} & 0\\
0 & 0 & 1
\end{array} }
\right].
\end{equation}
In the ($x$,$y$,$z$) basis, the stress is calculated via $\overline{\sigma} = \overline{\overline{R}}^\intercal\cdot\overline{\sigma'} \cdot \overline{\overline{R}}$ to be
\vspace*{-\baselineskip}
\begin{equation}
\label{eq:sigma}
\overline{\sigma_{m}}=\frac{\sigma}{2}
\left[ {\begin{array}{c}
1\\
1\\
0\\
0\\
0\\
1
\end{array} }
\right],
\end{equation}
where $\sigma$ is the magnitude of the stress applied. We can now apply Eq.\,\ref{eq:IndicatrixPhotoelasticReduced} to determine $\Delta \overline{B}$:
\begin{equation}
\label{eq:DeltaBm}
\Delta \overline{B_{m}}= \frac{\sigma}{2}
\left[ {\begin{array}{c}
q_{11} + q_{12}\\
q_{11} + q_{12}\\
2\,q_{12}\\
0\\
0\\
q_{44}\\
\end{array} }
\right].
\end{equation}
Since, however, we want to probe the birefringence in the ($x'$,$y'$,$z'$) basis, we apply the inverse rotation transformation ($\theta=\pi/4$) to determine $\Delta \overline{B'}$:
\begin{equation}
\label{eq:DeltaBprime}
\left[ {\begin{array}{c}
\Delta B'_{1}\\
\Delta B'_{2}\\
\Delta B'_{3}\\
\Delta B'_{4}\\
\Delta B'_{5}\\
\Delta B'_{6}\\
\end{array} }
\right]
=
\left[ {\begin{array}{c}
\Delta B'_{x'x'}\\
\Delta B'_{y'y'}\\
\Delta B'_{z'z'}\\
\Delta B'_{y'z'}\\
\Delta B'_{x'z'}\\
\Delta B'_{x'y'}\\
\end{array} }
\right]
= \frac{\sigma}{2}
\left[ {\begin{array}{c}
q_{11} + q_{12} + q_{44}\\
q_{11} + q_{12} - q_{44}\\
2\,q_{12}\\
0\\
0\\
0\\
\end{array} }
\right],
\end{equation}
from which follows (using Eq.\,\ref{eq:DeltaN}) a change in refractive index 
\begin{equation}
\label{eq:DeltaNprime}
\left[ {\begin{array}{c}
\Delta n_{x'x'}\\
\Delta n_{y'y'}\\
\Delta n_{z'z'}\\
\Delta n_{y'z'}\\
\Delta n_{x'z'}\\
\Delta n_{x'y'}\\
\end{array} }
\right]
=
\left[ {\begin{array}{c}
n_{x'x'}-n_0\\
n_{y'y'}-n_0\\
n_{z'z'}-n_0\\
n_{y'z'}-n_0\\
n_{x'z'}-n_0\\
n_{x'y'}-n_0\\
\end{array} }
\right]
= -\frac{\sigma}{4} \, n^3_0
\left[ {\begin{array}{c}
q_{11} + q_{12} + q_{44}\\
q_{11} + q_{12} - q_{44}\\
2\,q_{12}\\
0\\
0\\
0\\
\end{array} }
\right].
\end{equation}
We are primarily interested in the birefringence between axes ($x'$,$y'$), namely $\Delta n = \Delta n_{x'x'} - \Delta n_{y'y'} = n_{x'x'} - n_{y'y'}$. In the experiment, the stress is applied along $y'$. In this case, $\overline{\sigma'}=\left[0\,1 \,0 \,0 \,0 \,0 \right]^\intercal$, and $\Delta n = n_{x'x'} - n_{y'y'} = \frac{n^{3}_{0}}{2} \cdot q_{44} \cdot \sigma$, where $\sigma$ in this case has the inverse sign as in the case of stress applied along $x'$, from which we obtain Eq.\,\ref{eq:deltaN}.

\subsection{Bandgap shift with uniaxial stress}
In order to calculate the excitonic emission shift as a result of uniaxial stress (along $x'$) we assume that the shift is determined solely by the shift in the bandgap of the host material, GaAs. We start with the Bir-Pikur expression \cite{BirPikur1974} for the bandgap shift $\Delta E_{\rm gap}$ with applied strain $\overline{\overline{\epsilon_{ij}}}$ in the usual basis ($x$,$y$,$z$). The quantum dots themselves define the quantization axis, i.e.\ $z = [001]$. Assuming further that the valence state is of pure heavy-hole character, on account of the large heavy-hole--light-hole splitting, 
\begin{equation}
\label{eq:BirPikur}
\Delta E_{\rm gap} = a_{\Gamma} \, {\rm Tr}\left(\epsilon_{ij}\right) + \frac{b}{2 \, }(2\cdot\epsilon_{zz} - \epsilon_{xx} - \epsilon_{yy}),
\end{equation}

\noindent where $a_{\Gamma}$ and $b$ are the deformation potential coefficients, and ${\rm Tr}(\epsilon_{ij})$ is the trace of the strain tensor $\overline{\overline{\epsilon_{ij}}}$.

We apply a stress, and thereby induce a strain. We use the strain--stress relation $\overline{\epsilon_m} = \overline{\overline{S_{mn}}} \cdot \overline{\sigma_n}$, where $\overline{\overline{S_{mn}}}$ is the compliance tensor, abbreviated in a similar way to Eq.\,\ref{eq:IndicatrixPhotoelasticReduced} on account of symmetry. We now know also the expression for a uniaxial stress along $x'$ in the usual basis (Eq.\,\ref{eq:sigma}). The strain--stress relation reads
\begin{equation}
\begin{aligned}
\label{eq:strain-stress}
\left[ {\begin{array}{c}
\epsilon_{xx}\\
\epsilon_{yy}\\
\epsilon_{zz}\\
\epsilon_{yz}\\
\epsilon_{xz}\\
\epsilon_{xy}\\
\end{array} }
\right]
& = \frac{\sigma}{2} \,
\left[ {\begin{array}{cccccc}
S_{11} & S_{12} & S_{12} & 0 & 0 & 0\\
S_{12} & S_{11} & S_{12} & 0 & 0 & 0\\
S_{12} & S_{12} & S_{11} & 0 & 0 & 0\\
0 & 0 & 0 & S_{44} & 0 & 0\\
0 & 0 & 0 & 0 & S_{44} & 0\\
0 & 0 & 0 & 0 & 0 & S_{44}\\
\end{array} }
\right]
\left[ {\begin{array}{c}
1\\
1\\
0\\
0\\
0\\
1\\
\end{array} }
\right] \\
& = \frac{\sigma}{2} \,
\left[ {\begin{array}{c}
S_{11} + S_{12}\\
S_{11} + S_{12}\\
2 \, S_{12}\\
0\\
0\\
S_{44}\\
\end{array} }
\right].
\end{aligned}
\end{equation}

Now, ${\rm Tr}(\epsilon_m) = \epsilon_{xx} + \epsilon_{yy} + \epsilon_{zz} = (S_{11}+2\,S_{12})\,\sigma$ and $2\,\epsilon_{zz} - \epsilon_{xx} - \epsilon_{yy} = (S_{12}-S_{11})\,\sigma$. Transforming\cite{Nye1957} $\overline{\overline{S_{mn}}}$ to the stiffness tensor $\overline{\overline{C_{mn}}}$, Eq.\,\ref{eq:BirPikur} gives us the bandgap shift as a function of a stress along $[110]$:
\begin{equation}
\label{eq:BandGapshiftStress}
\delta E_{\rm gap}/\delta\sigma = \left(\frac{a_{\Gamma}}{C_{11}+2\,C_{12}} - \frac{b}{2} \, \frac{1}{C_{11}-C_{12}}\right).
\end{equation}

We use the following numerical values for GaAs from literature \cite{Burenkov1973,VandeWalle1871,Sun2007} for the computation: $a_{\Gamma}=-8.33$\,eV, $b = -2.00$\,eV, $C_{11}=122.3$\,GPa and $C_{12}=57.1$\,GPa. We finally arrive at $\delta E_{\rm gap}/\delta\sigma=-22.2\,\mu$eV/MPa.

\subsection{Piezo-optical coefficients $q_{44}$ at T = 4\,K}

Data on the piezo-optical coefficient $q_{44}$ of Al$_x$Ga$_{1-x}$As alloys can be found for measurements \cite{Feldman1968,Adachi1985} at $T = 298$\,K, and for GaAs at $T = 77$\,K \cite{Feldman1968}. However, this data is not available at $T = 4$\,K to the best of our knowledge. The dispersion of these coefficients is linked to the bandgap of the particular material. In particular, $q_{44}$ shows a resonance behaviour at the bandgap itself. As the bandgap of these materials shifts with temperature, the $q_{44}$ coefficients are temperature dependent. It is therefore necessary to estimate the $q_{44}$ values at $T = 4$\,K. We elaborate here the procedure.

\begin{figure}[t!]
\includegraphics[width=\columnwidth]{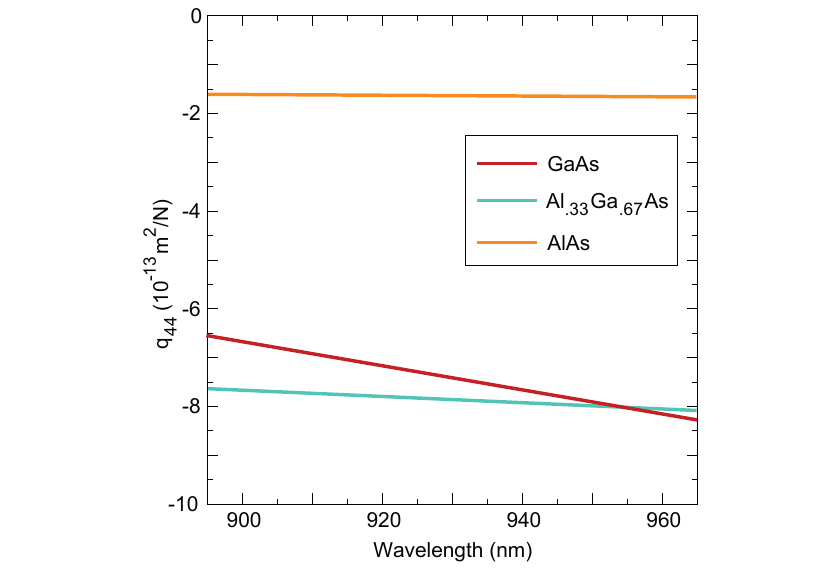}
\caption{\label{fig:q44} Piezobirefringent coefficients $q_{44}$ at $T = 4$\,K for GaAs, Al$_{.33}$Ga$_{.67}$As and AlAs, estimated from their room temperature values by shifting the wavelength rigidly by an amount equal to the shift in bandgap with temperature.}
\end{figure}

Adachi \cite{Adachi1985} provides data -- we extract the data from the plots with Webplotdigitizer \cite{Rohatgi2020} -- on Al$_x$Ga$_{1-x}$As alloys, of particular relevance here the dispersion curve of the elasto optic coefficients $p_{44}$, related to the piezo-birefringent coefficients via $q_{44}=p_{44} \cdot S_{44}$. The room-temperature bandgap energies of the alloys of interest are also extracted ($E_{\rm gap}$(GaAs)$=1.424$\,eV, $E_{\rm gap}$(Al$_{.33}$Ga$_{.67}$As)$=1.8355$\,eV, $E_{\rm gap}$(AlAs)$=2.168$\,eV). The optical properties of semiconductor crystals, such as the refractive index, are linked to the bandgap energy of the material \cite{Navindra2007}. The temperature dependence of the GaAs bandgap can be described via $E_{\rm gap}(T) = E_{\rm gap}(0) - 5.405 \cdot 10^{-4} T^2/ \left(T + 204\right)$ (with $E_{\rm gap}(T)$ in eV, $T$ in K) \cite{Blakemore1982}. This equation was demonstrated to be valid also for Al$_x$Ga$_{1-x}$As alloys \cite{Lourenco2001}.

From the room-temperature (298\,K) bandgap energies, we can estimate the low-temperature bandgap energies of our materials, namely $E_{\rm gap}$(GaAs)$=1.519$\,eV, $E_{\rm gap}$(Al$_{.33}$Ga$_{.67}$As)$=1.931$\,eV, $E_{\rm gap}$(AlAs)$=2.263$\,eV, representing a shift in bandgap energy of 95\,meV for these materials. These shifts translate into a shift in wavelength of $\Delta\lambda=-54.45$\,nm, $\Delta\lambda=-33.22$\,nm and $\Delta\lambda=-24.01$\,nm, respectively. We now estimate $q_{44}$ at 4\,K for a particular wavelength $\lambda$ by rigidly shifting the curve of $q_{44}$ versus $\lambda$ at 298\,K by $\Delta\lambda$. We confirm that this method functions well by comparing translated $T = 298$\,K data\cite{Adachi1985} for $q_{44}$ to $T = 77$\,K data \cite{Feldman1968} and verifying an overlap.

Finally, we comment that the dispersion of $q_{44}$ of the semiconductor materials is rather small in the spectral band of interest, as exemplified in Fig.\,\ref{fig:q44}, such that we use their mean values in the model -- we treat the small dispersion as a measure of the uncertainty in the parameters.

\begin{acknowledgments}
We thank Liang Zhai for helpful advice on the experimental setup. We acknowledge financial support from SNF project 200020\_175748, NCCR QSIT and Horizon-2020 FET-Open Project QLUSTER. A.J.\ acknowledges support from the European Unions Horizon 2020 Research and Innovation Programme under the Marie Sk{\l}odowska-Curie grant agreement No.\ 840453 (HiFig). S.R.V., R.S., A.L.\ and A.D.W.\ acknowledge gratefully support from DFH/UFA CDFA05-06, DFG TRR160, DFG project 383065199, and BMBF Q.Link.X No. 16KIS0867.\\
\end{acknowledgments}

\bibliography{PhotoelasticityCavity_v3.1}

\end{document}